# Hunting Structural Demons in Digital Reticular Chemistry: Lessons from Metal-Organic Frameworks

Yongchul G. Chung*[a], Myoung Soo Lah[b]

*Dedicated to Professor Omar Yaghi on the occasion of the 2025 Nobel Prize in Chemistry*

**Abstract:** Digital reticular chemistry relies on accurate crystal structures to power computational screening, data-driven discovery, and structure-property analysis, yet recent studies reveal that more than half of the top-performing candidates in major computational screening campaigns are chemically invalid. In experimental MOF databases, structural errors arise when disordered or incomplete structural models are incorrectly converted into fully specified simulation inputs. In hypothetical MOF database, structures are complete by construction but may encode chemically implausible oxidation states, coordination environments, or charge distributions. We term these erroneous structural models "structural demons." This mini-review asks three questions: where these errors enter, how we find them, and how we prevent them. On the prevention side, the key steps are keeping diffraction data and synthesis details together from the start, using consistent curation when structures enter a database, and filtering topology choices before structure generation. Connecting these steps can keep many bad structures out of downstream databases and reduce the need to fix them later.

## 1. Introduction

Reticular chemistry, defined as the construction of extended networks from molecular building units joined by strong bonds, offers a virtually limitless design space of porous materials[1]. Following the seminal reports of porous metal-organic frameworks (MOFs) in the 1990s[2], Yaghi and coworkers formalized the concept of reticular synthesis in 2003[3], demonstrating that careful selection of secondary building units (SBUs) could yield porous networks with predetermined structures and properties. Since then, tens of thousands of unique framework materials have been synthesized and compiled in the Cambridge Structural Database (CSD)[4] and elsewhere[5], yet this represents only a tiny fraction of the potential design space. Building on this foundational principle, computational approaches have generated hundreds of thousands of MOFs[6] and covalent-organic frameworks (COFs)[7] *in silico*, creating vast libraries that can be computationally screened for applications ranging from carbon capture[8] and energy storage[9] to water harvesting[10] and catalysis[11]. Yet this abundance of structures raises a question that is arguably as important as finding optimal candidates: how reliable are the crystal structures that underpin the entire computational screening pipeline?

High-throughput computational screening is a powerful approach to navigate through large collections of MOF structures and identify candidates with optimal properties for experimental synthesis and testing. A fundamental assumption in these studies is that the input structural models are fully specified and chemically consistent enough to run molecular simulation, often referred to as "computation-ready" (CR) structures. Unfortunately, this assumption is frequently violated. Several recent studies have revealed that a significant fraction of structural models in MOF databases, both experimental and hypothetical, contain errors: missing or misplaced atoms, incorrect bond connectivity, and unbalanced framework charges[12]. Even a single misassigned proton can change a framework's formal charge, metal oxidation state, and predicted adsorption performance in molecular simulation. White et al. demonstrated that when top-performing candidates from eight recent high-throughput screening studies were re-examined, 52% were found to be chemically invalid[12b]. We refer to these erroneous structural models as "structural demons" throughout this review. These demons do more than distort individual screening studies. They can bias structure-property relationships, contaminate machine-learning (ML) training sets, and misdirect experimental prioritization by making chemically implausible frameworks appear competitive. As hypothetical databases increasingly serve as pre-training data for predictive and generative models, errors in the data become errors in the models, making database quality an issue that extends well beyond curation alone. Recognizing how widespread these problems are, the community has developed a growing arsenal of methods[12a, 12c, 12d, 13] to hunt these demons, from rule-based validators and ML classifiers to standardized data formats that aim to prevent new demons from being introduced in the first place.

[a] *Yongchul G. Chung*
School of Chemical Engineering, School of Transdisciplinary Engineering, AI Convergence Computational Science, and Graduate School of Data Science, Pusan National University
Busan, Korea (South) 46241
phone: +82 51 510 3757
e-mail: drygchung@pusan.ac.kr

[b] *Myoung Soo Lah*
Department of Chemistry, Ulsan National Institute of Science and Technology, Ulsan, Korea 44919

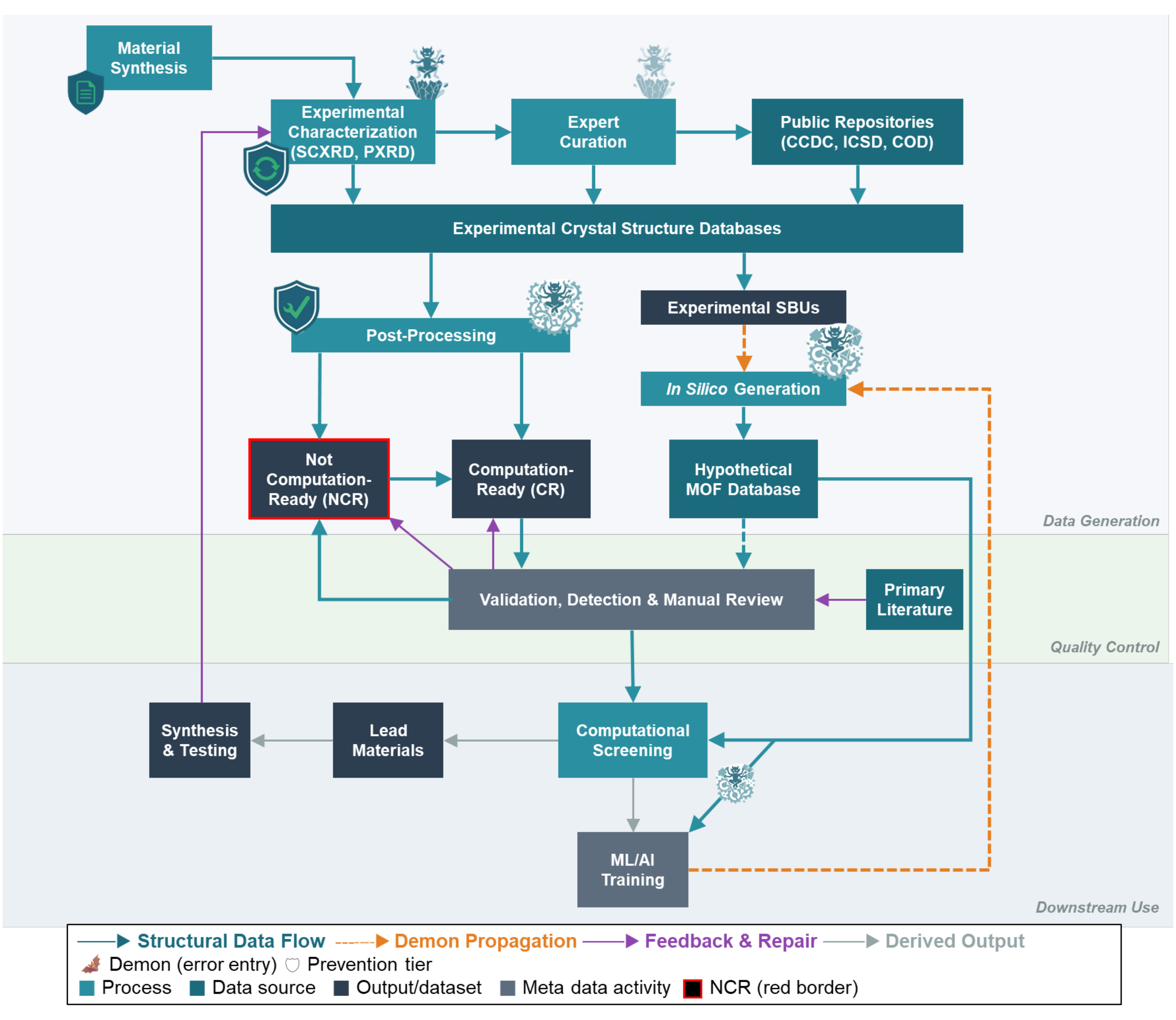

**Figure 1.** The digital reticular chemistry pipeline, showing how structural demons enter at four stages and propagate downstream. The pipeline spans data generation (top), quality control (middle), and downstream use (bottom). Demon icons mark the four error entry points: experimental characterization (**D1**), expert curation (**D4**), automated post-processing (**D2**), *and in silico* generation (**D3**). Orange dashed arrows trace propagation pathways, including contaminated SBU inheritance and the model-native feedback loop through ML/AI training. The dashed teal arrow from the hypothetical MOF database to validation indicates that systematic chemical screening of hypothetical libraries is recommended but not yet routine. Purple arrows denote feedback and repair, including the experimental validation cycle from synthesis back to characterization. Shield icons mark the three prevention tiers (**Section 4**). Gray arrows indicate derived outputs.

The review is organized in three parts. First, we examine how crystal structures enter the digital reticular chemistry workflow (**Figure 1**), with emphasis on the different demons that arise in experimental and hypothetical databases. Second, we discuss how those demons are detected and classified using rule-based methods, ML models, and literature-grounded systems. Third, we consider how data standardization and integrated infrastructure can reduce the introduction of new demons while strengthening the computational foundation for reticular material discovery. This review does not simply list databases and tools. Instead, it follows three linked questions: where the errors enter, how we find them, and how we stop them. The second question also includes how the errors spread once they enter the pipeline. Although the broader theme is digital reticular chemistry, used here in a data-centric sense, encompassing structure generation, curation, and validation, and digital interfaces to characterization, the discussion below focuses mainly on MOFs, with COF examples included when they clarify general lessons about validation, data infrastructure, or experimental standardization. A recurring theme is the absence of a centralized community-governed structure repository for reticular materials, a gap that shapes many of the challenges discussed below.

## 2. Crystal Structural Data Generation and Collection

The ecosystem of digital reticular chemistry draws on crystal structures from two fundamentally distinct sources: experimental structure data collections and hypothetical or generated databases. Although these are often combined in computational studies, their origins, metadata, and dominant limitations differ. Making this distinction explicit is essential for understanding how structural demons arise and how they should be interpreted.

### 2.1. Experimental Structure Collections

Tens of thousands of MOF structures have been reported in the literature and deposited in crystallographic databases, most notably the CSD, which contains more than 100,000 MOF-like entries[4b, 5b]. Converting these raw crystallographic files into forms suitable for molecular simulation has been a defining challenge of digital reticular chemistry, and the Computation-Ready Experimental MOF (CoRE MOF) database was one of the first systematic efforts to address it[5a, 14]. Introduced in 2014 as part of the U.S. Materials Genome Initiative, CoRE MOF applied automated workflows to remove solvent molecules, standardize atom labels, and convert experimental CIFs into computation-ready form, initially curating approximately 4,700 porous 3D structures[5a]. Sholl et al. performed DFT-optimization and assigned density-derived electrostatic and chemical (DDEC) partial atomic charges[15] on this database (CoRE MOF-DFT[16] and CoRE MOF-DDEC[17]). A major update in 2019 expanded the database to more than 14,000 structures and introduced improved procedures for disorder handling, hydrogen restoration, and structural analysis[5c].

More recently, CoRE MOF database (CoRE MOF DB) evolved into a feature-rich, screening-ready resource containing over 40,000 experimental MOFs reported by early 2024[5d]. In addition to curated crystal structures, the database provides geometric descriptors from Zeo++[18], topology assignments matched to the RCSR database[19], building-block decomposition through an updated MOFid workflow[20], machine-learned DDEC6 partial charges, and ML-predicted stability-related properties[21]. These additions make the dataset more useful for integrated material-process screening[5d, 22] and large-scale ML studies.

A key change in the newer CoRE MOF releases is that the updated workflow distinguishes CR from not-computation-ready (NCR) structures rather than forcing every entry into a chemically completed model. The 2025 report[5d] describes 17,202 CR structures and 23,635 NCR structures, showing that a substantial fraction of entries remain in NCR status. This is best understood as a conservative curation choice: preserving the original crystallographic information is often safer than applying aggressive automated corrections that could introduce new structural demons. A later workflow[12d] update also added MOFClassifier as an additional validation layer.

While CoRE MOF DB has served as the main computation-ready resource for more than a decade, several parallel efforts have taken complementary approaches to organizing and enriching experimental structural data. Moghadam et al. developed the CSD MOF Subset in 2017 in collaboration with the Cambridge Crystallographic Data Centre (CCDC), taking a fundamentally different approach[5b]. Rather than converting CIFs into computation-ready form, this effort focused on identifying which CSD entries are MOFs in the first place, applying seven chemical bonding criteria and providing solvent-removal scripts for on-demand activation. The subset is automatically updated with each CSD release and was later extended into the freely accessible CSD MOF Collection, which currently provides over 12,000 three-dimensional porous MOFs for academic research. Burner et al. addressed a different gap with ARC-MOF[23], which spans approximately 280,000 MOFs from all publicly available sources and provides DFT-derived REPEAT charges[24] for each structure. Other notable efforts include the QMOF database[25], which subjects structures to DFT-level optimization and strict validation before computing quantum-chemical properties, and the CURATED COFs database[7a], which applies similar curation principles to covalent organic frameworks. Each of these databases addresses a different stage of the pipeline from raw crystallographic data to simulation-ready input, and together they form an interconnected ecosystem in which the same underlying CSD or literature sources may pass through multiple curation and enrichment workflows.

A common thread across all these databases is that the underlying crystal structures originate from single-crystal X-ray diffraction (SCXRD) or powder X-ray diffraction (PXRD) experiments, which determine a time- and space-averaged electron density map. Crystallographic models therefore commonly include disorder with partial occupancies, symmetry-constrained split atomic positions, unresolved hydrogen atoms, and diffuse solvent or counter-ion contributions. These conventions are chemically meaningful within the crystallographic context, but converting them into fully ordered, single-configurational, charge-balanced representations with properly assigned hydrogen atoms for molecular simulation inevitably requires chemical judgment that is difficult to automate reliably. It is at this interface between crystallographic convention and computational requirement that structural demons readily enter the digital ecosystem.

### 2.2 Hypothetical Structure Generation

Hypothetical MOF generation has been a central pillar of digital reticular chemistry, enabling systematic exploration of framework space beyond experimentally synthesized materials. Because these structures are assembled *in silico* from predefined building blocks and algorithms, they are generated as single-configuration structural models. Accordingly, hypothetical MOFs generally do not exhibit disorder, partial occupancies, or symmetry-averaged atomic positions that often render experimental CIFs not computation-ready. Their dominant failure mode is fundamentally different: hypothetical structures are usually complete at the file level, but may still encode chemically implausible coordination environments, oxidation states, or charge distributions. The extent to which this failure mode can be avoided depends critically on whether chemical feasibility is enforced at the topology selection stage or verified after structures are fully generated. We review the history of hypothetical MOF generation algorithms to show when and why the structural demons are created.

### 2.2.1 Bottom-up Enumeration

Early large-scale hypothetical MOF databases were constructed using bottom-up enumeration, in which inorganic nodes and organic linkers are assembled according to local connectivity rules without using any topological information. The landmark work by Wilmer et al.[6a] used this approach: from a library of 102 building blocks derived from the crystallographic data of existing MOFs, they generated 137,953 hypothetical structures by stepwise recombination of building blocks at their connection sites without any force-field energy minimization. The method was validated by reconstructing known MOFs (including IRMOF-1, HKUST-1, PCN-14, and MIL-47) as hypothetical analogues and confirming that atomic positions agreed to within 0.1 angstrom on average. A three-stage methane storage screening then identified over 300 hypothetical candidates predicted to exceed the best experimentally known storage capacity at the time. The same hMOF library was then applied to $CO_2$ separation and capture, showing that optimal adsorbent characteristics depend on separation conditions[8a]. Moreover, only six topologies (**pcu**, **dia**, **nbo**, **fcu**, **tbo, sra**) arose naturally from the assembly rules rather than being specified as an input, with 90% of the nets being 6-coordinated **pcu**[26]. Because topology is an emergent outcome rather than a controlled variable, bottom-up enumeration makes it difficult to disentangle the roles of building-block connectivity and symmetry in determining framework properties. The strong bias toward a 6-coordinated **pcu** net also means that chemical diversity across different coordination environments remains underexplored in these databases.

### 2.2.2 Top-down, topology-guided crystal completion

To overcome these limitations, later hypothetical MOF generators adopted top-down, topology-guided strategies, in which the target topology is fixed *a priori* as a periodic abstract net and the computational task is to map topologically compatible building blocks onto that blueprint. The reversed topological approach (RTA) introduced by Schmid and co-workers in 2015 generated multiple isoreticular isomers of (4,4)-c **nbo-b** topology by combining square-planar 4-c copper paddle-wheel nodes with tetracarboxylate linkers[27], and Boyd and Woo in 2016 developed a graph-theoretical method for constructing MOFs using predefined SBUs and a target 3D net topology represented by a labelled quotient graph[28]. By exploiting the fact that a single net can have infinitely many embeddings in 3D space, the method adjusts these embeddings so that the net's vertices align with the geometries of the chosen SBUs. The SBUs placed at node positions were subsequently optimized using classical force field parameters. The approach was demonstrated by generating MOFs from 46 different network topologies using the same pair of 3- and 4-coordinated SBUs. This method enables the creation of large, topologically diverse hypothetical MOF databases for computational screening. Top down crystal completion provides internally consistent reference structures for molecular simulation workflows and enables systematic comparison of frameworks with different building blocks sharing the same connectivity. An important consequence is that topology-property relationships can be investigated varying only the identity of building blocks while keeping the underlying net topology. This shows how the building blocks alone govern properties such as pore geometry, surface areas, and mechanical properties. For example, Haranczyk and colleagues demonstrated that decorating a single topology (such as the MOF-74 **rod** net) with a series of linkers and metals can isolate topology-controlled adsorption and transport behaviour[29].

Colón, Gómez-Gualdrón, and Snurr provided the first detailed, publicly available implementation of a topology-based crystal constructor, ToBaCCo, which reads a topological blueprint as input, identifies compatible building blocks based on coordination and symmetry, and assembles the periodic crystal structure accordingly[6b]. Using ToBaCCo, the authors constructed and screened 13,512 MOFs across 41 edge-transitive topologies for hydrogen storage and noble-gas separation, demonstrating strong topology-dependence across 41 nets. Anderson and Gómez-Gualdrón[6c] later introduced ToBaCCo 3.0 with a crystallographic net rescaling algorithm that expanded coverage to previously inaccessible topologies, showing that methane deliverable capacity varies up to 150 cc(STP)/cc across structures built from the same building blocks but assembled in different topologies. Boyd et al. applied a related topology-based constructor, TOBASCCO, to generate a library of over 325,000 hypothetical MOFs spanning multiple topologies and used high-throughput screening to identify a common $CO_2$-binding motif, termed an "adsorbaphore," that maintains selectivity even under humid flue-gas conditions[30]. Two predicted materials were subsequently synthesized and experimentally confirmed to perform as the simulations predicted, providing one of the most complete demonstrations to date of the pipeline from topology-guided generation through computational screening to experimental validation. PORMAKE[6d] took a complementary approach by treating the target pore architecture, rather than the net topology alone, as the primary design variable. The code accepts any user-defined topology as input, constructs the corresponding periodic crystal by placing building blocks at node and edge positions, and evaluates geometric feasibility through overlap and distance checks. By tabulating the geometric properties of individual building blocks and topologies independently, trillions of hypothetical building block/topology combinations could be screened without explicitly constructing each structure, enabling rapid identification of promising candidates for methane storage from a large combinatorial space.

A conceptually important development was recently introduced by Darù et al.[31], who proposed a symmetry-guided topology filtering workflow as a preprocessing step for top-down generation. Rather than optimizing every topologically compatible structure, the workflow uses the point-group symmetry of vertices (nodes) and edges (linkers) to filter an initial topology list derived solely from connectivity. Symmetry groups of vertices and edges at adjacent levels (i.e., subgroups and supergroups) are also examined to account for the slight distortions that occur when building blocks are assembled into a periodic framework. The surviving candidates are ranked by occurrence frequency in the RCSR database, and the top-ranked subset is passed to ToBaCCo or TOBASCCO for CIF generation, followed by DFT optimization. The workflow was validated by blind reconstruction of ten known MOF and COF structures; in the case of MOF-841, symmetry filtering reduced fifteen topology candidates to a single correct assignment (**flu**). The method does not yet handle interpenetration and currently requires downstream tools for CIF file generation. The implication of this approach for reducing chemical invalidity in hypothetical databases are discussed in **Section 2.2.3.**

All current generators, whether bottom-up or topology-guided, assemble crystal structures from discrete building blocks; they differ primarily in how much topological and symmetry information is used during that assembly, and how these design choices produce chemical invalidity is the subject of the next subsection.

### 2.2.3 Chemical Invalidity of Hypothetical MOFs and Its Origins in Current Generators

Chemical invalidity in hypothetical MOFs is not a single failure mechanism but a family of mechanisms that can arise before, during, or after structural model assembly. Two sources of error are particularly important: SBU inputs carrying incorrect atomic positions, missing atoms, or distorted coordination geometries from their source structures, and insufficient use of topological and symmetry constraints during generation. Because many high-error hMOF datasets were built from CoRE-derived SBUs using their own generation algorithms, these two sources are entangled: a high structural error rate in a given database cannot be attributed to SBU quality or algorithm design alone without controlled comparisons that hold one factor fixed while varying the other. Gibaldi et al. curated a validated SBU library (HEALED)[32] precisely to reduce the input side of this problem.

On the generator side, the topological and symmetry information encoded in the RCSR database[19] is still used sparingly, if at all, during structure generation. For each net topology, the RCSR code records the space group, lattice parameters of the maximum-symmetry embedding with unit edge length of a net, the coordination numbers, vertex and edge symmetries, the barycentric-embedded coordinates. These descriptors restrict compatible net topologies based on the connectivities and symmetries of any given building blocks, substantially reducing the number of periodic networks that needs to be completed in three-dimensional space. Most existing generators use only a subset of this information, and the consequence is an unnecessarily large candidate space populated by chemically unreasonable structures.

Bottom-up enumeration sidesteps topology altogether. As noted in Section 2.2.1, the Wilmer database is dominated by a single six-coordinated net (**pcu**), leaving chemical diversity across other coordination environments largely unexplored. In the original ToBaCCo implementation[6b], building-block compatibility was determined by matching the number of connection points to the coordination number of the net vertices, together with an integer symmetry code that distinguishes, for example, tetrahedral ($T_d$) from square-planar ($D_{4h}$) four-coordinated nodes. The net itself was treated in space group P1 regardless of its true symmetry, and only isotropic scaling was available to adjust the blueprint to the building-block geometry. Anderson and Gómez-Gualdrón introduced an anisotropic rescaling algorithm built on the voltage-graph (quotient-graph) representation of the net, following the formalism of Eon[33], which expanded the coverage of 19 previously inaccessible non-edge-transitive topologies. Even so, the code still operates in *P*1 and does not enforce the space-group symmetry of the target net as a construction constraint.

The restriction to edge-transitive nets in the original ToBaCCo is a computational convenience: because all edges in such nets are symmetry-equivalent, the unit cell can be resized with a single isotropic scaling factor. Non-edge-transitive nets require anisotropic rescaling, which is more complex but necessary for the majority of experimentally observed topologies. Many of the 41 edge-transitive nets used as ToBaCCo blueprints have rarely or never been observed experimentally. As the topology frequency analysis of Darù et al.[31] demonstrates, only a handful of nets (notably **pcu**, **nbo**, **pts**, **tbo**) dominate both hypothetical and experimental databases; directing computational resources toward nets with demonstrated experimental viability, rather than exhaustively enumerating all possible edge-transitive nets, would increase the likelihood that the generated structures correspond to synthesizable materials.

These limitations have measurable consequences. White et al. evaluated over 1.9 million structures across 14 major experimental and hypothetical MOF databases and found error rates exceeding 40% in most hypothetical libraries[12b]. Even databases not derived from CoRE SBUs, such as the Wilmer (40%) and Boyd-Woo (21%) libraries, carry substantial error rates. When these structures serve as pre-training data for ML models [34] and generative models[35], the errors propagate into the models themselves: property prediction models may assign artificially favorable properties to chemically invalid structures, and generative models may produce new outputs that inherit the same implausibilities, creating a feedback loop that amplifies structural demons rather than eliminating them.

The structural demons in hypothetical MOF databases therefore reflect both contaminated SBU libraries and the limited extent to which current generators encode the topological and symmetry information already available in RCSR. Future generation workflows will need to incorporate this information more aggressively, using vertex and edge site symmetry as construction-time filters rather than post-hoc descriptors, if the fraction of chemically unreasonable structures entering hypothetical databases is to be substantially reduced.

In summary, experimental and hypothetical MOF databases include structural demons of different kinds. In experimental databases, structural demons arise from a gap between the crystallographic model and the fully specified representation required for molecular simulation. In hypothetical databases, structural demons take the opposite form: the structures are complete by construction yet chemically implausible. These different origins demand different diagnostic and curation strategies, which is the subject of the next two sections.

## 3. Origins, Detection, and Classification of Structural Demons

The structural demons discussed here fall into two related but distinct failure classes. First, some structures are NCR: a unique simulation model cannot be specified without additional assumptions. Second, others are CR yet chemically invalid because the completed structural model itself is implausible. Experimental databases contain both classes, whereas hypothetical databases are dominated by the second. This distinction matters because the appropriate response depends on the failure class: NCR structures may require reinterpretation of raw crystallographic data through rule-based methods and/or consultation of the literature, whereas chemically invalid but algorithmically complete structures require chemical validation and where possible, repair.

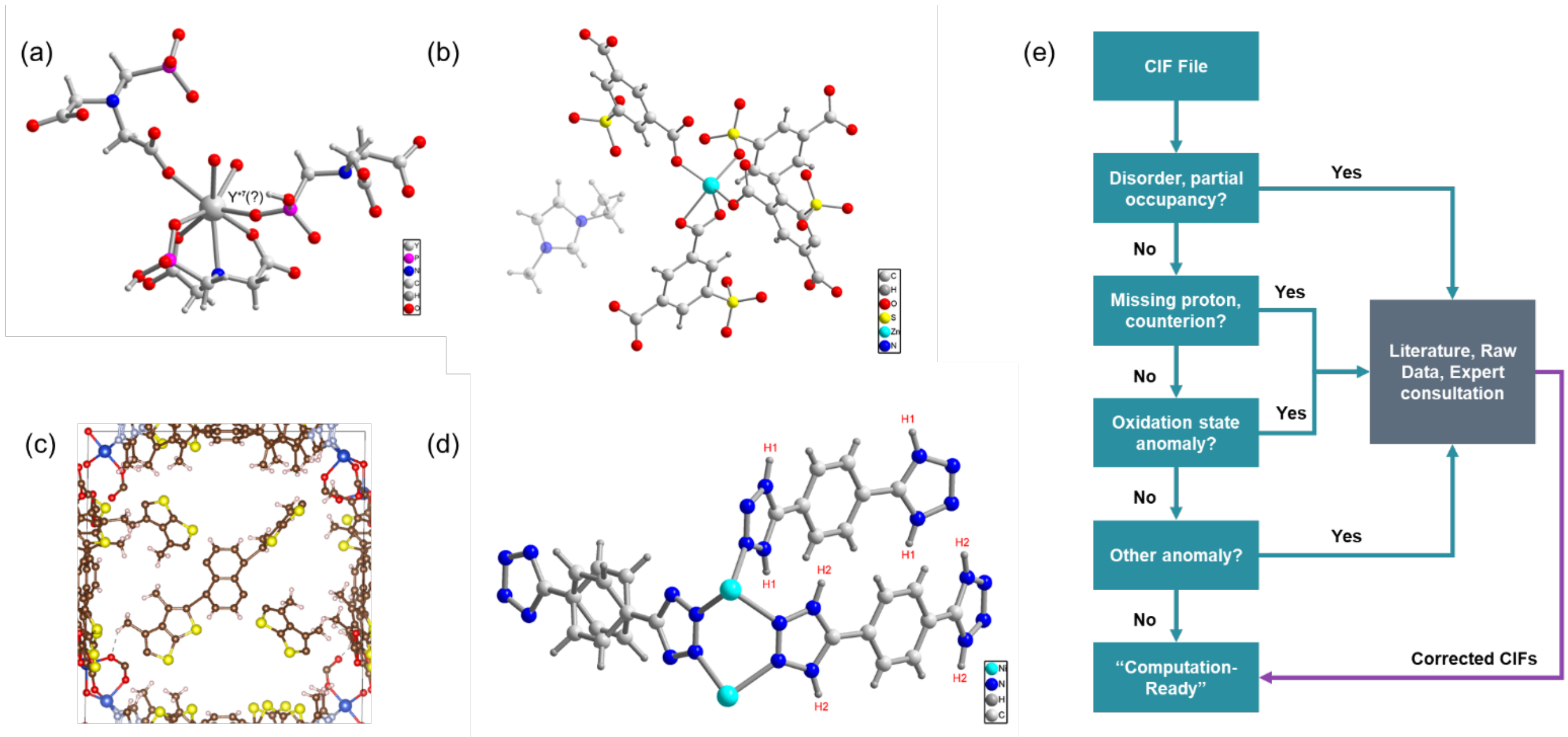

**Figure 2.** Examples of structural demons from experimental MOF databases and a decision framework for classifying them. (a) SEYDUW (**D1**): missing hydrogen atoms cause aqua ligands to appear as oxo ligands, yielding an impossible $Y^{7+}$ oxidation state. (b) DUNXUH (**D2**): automated solvent removal omits charge-balancing imidazolium cations, making Zn appear 3+ rather than 2+. (c) hMOF from PORMAKE (**D3**): a hypothetical MOF with ftw topology, generated using 4-connected N564, 12-connected N619, and 2-connected E160 using PORMAKE and flagged by MOSAEC workflow. The overall valency of the structure is not charge neutral. (d) BAKGIF (**D4**): incorrect proton reassignment during curation produces an erroneous protonation state on ligand. (e) Decision tree for identifying and routing structural errors.

### 3.1 Crystallographic Origins of Structural Demons

The demon entry points identified in **Figure 1** can be grouped into four categories by the stage at which they enter the pipeline. **Figure 2** is most useful when read as a casebook for these entry points: each panel should show the source state, the step at which the demon is introduced, and the resulting chemically invalid outcome. Making this classification explicit is essential because the appropriate intervention differs in each case.

**(D1) Experimental characterization.** The first and most fundamental source of structural demons lies in the crystallographic model itself. X-ray diffraction determines a time- and space-averaged electron density map; therefore, deposited CIF files routinely contain disordered molecular fragments, split atomic positions with partial occupancies at crystallographic symmetry sites, missing and/or misassigned hydrogen atoms, diffuse solvent and/or counterion contributions, which are often handled by the SQUEEZE/PLATON procedure. These conventions are standard for structure reporting but become problematic when a unique, fully ordered atomic model is required for molecular simulation. The critical distinction is whether a given species is framework-defining (such as a charge-balancing counterion, coordinated solvent, or an organic/inorganic ligand with an appropriate protonation state) or a genuinely removable pore guest. Because hydrogen scatters X-rays weakly[36], they are often missing or misplaced in deposited CIF files, although the presence or absence of a single hydrogen atom can change the formal charge, assigned metal oxidation state, and predicted adsorption behaviour. A representative example is SEYDUW (**Figure 2a**), where the hydrogen atoms on two coordinated solvent water molecules are missing due to the limited sensitivity of X-ray diffraction to hydrogen scatterers[12b]; the original CSD entry, with two oxo ligands, misassigns yttrium an impossible 7+ oxidation state, whereas the source publication reports Y(III) with aqua ligands. The protonation state of ligated oxygen atoms can be assigned based on a combination of the coordination geometry such as metal-oxygen distance and the probable oxidation state of the metal, or by consulting the primary literature.

**(D2) Automated post-processing.** The second major source of demons arises during the automated workflows that convert raw CIF files into computation-ready representations.

Solvent removal, disorder resolution, hydrogen addition, and counterion handling require chemical judgment that is difficult to encode fully in a generic pipeline, and errors at any step can propagate downstream. Two technical choices in earlier workflows are especially prone to introducing such errors. First, the adjacency matrix that determines which atoms belong to the framework was built using a fixed skin distance (typically 0.25 Å added to the sum of covalent radii), and all molecular clusters smaller than the largest connected component were removed. This works for simple organic solvents, but it can misclassify metal ions that are weakly bound with ligands or large counterions as disconnected fragments, stripping them from the structure and breaking charge balance. The CoRE MOF 2024 workflow addresses this by replacing the fixed skin distance with a variable threshold that increases incrementally until every metal atom is included in the largest connected cluster, preventing accidental removal of framework metals. Second, solvent and ion removal typically relies on a predefined list of known molecular species matched by stoichiometry. When a listed ion is disordered in the crystal structure, its apparent atom count no longer matches the expected formula, so the algorithm fails to recognize it and removes it without checking whether it is charge-balancing. A representative case is DUNXUH (**Figure 2b**), where the computation-ready representation omits the charge-balancing imidazolium cations reported in the source publication, making Zn appear formally 3+ rather than 2+[12b]. Both problems share a common lesson: generic rules written for well-ordered structures break down precisely where crystallographic ambiguity is greatest. The SAMOSA protocol takes a complementary approach by monitoring metal oxidation states throughout the solvent-removal process, and flagging structures where the assigned state is chemically implausible [37].

The distinction between **D1** and **D2** matters in practice. **D1** demons require going back to the diffraction data or the original paper, whereas **D2** demons are fixable by improving the post-processing algorithm. The CoRE MOF 2024 workflow takes the conservative route of distinguishing CR from NCR structures rather than forcing every entry into a completed model, a strategy that avoids introducing new **D2** demons at the cost of leaving ambiguous entries unresolved.

**(D3) *In silico* generation.** Hypothetical MOF structures do not exhibit the disorder and partial-occupancy problems of experimental CIFs because they are algorithmically assembled as single-configuration models. Their demons take a different form: coordination environments and oxidation states that no real material would adopt. These errors arise from at least three sub-mechanisms.

The first is contaminated input. When SBU libraries are extracted from experimental databases that already contain **D1** or **D2** errors, the extracted building blocks may carry incorrect atomic positions, missing atoms, or distorted coordination geometries that reflect the errors in the source structure. Gibaldi et al. built the HEALED library[32] specifically to break this chain by validating each SBU before it enters the generation pipeline.

The second is topology-chemistry mismatch. As discussed in **Section 2.2.3**, most generators check only coordination number when placing an SBU on a net, without verifying whether the SBU's point-group symmetry is actually compatible with the vertex site symmetry of that net. The result can be a structure where, for instance, a tetrahedral SBU is placed on a vertex whose site symmetry requires square-planar coordination, producing a framework with unrealistic linker connection angles and steric strain at the node. **Figure 2c** illustrates the second sub-mechanism. The hypothetical MOF is assembled from three building blocks (a naphthalene dicarboxylate linker, a thieno[3,2-b]thiophene linker, and a Cu-O-Cu bridging SBU) whose coordination numbers match the target net, but whose point-group symmetries are not verified against the vertex and edge site symmetries of that net. The resulting structure is algorithmically complete yet carries a net negative framework charge, indicating that the formal charges of the chosen building blocks do not sum to zero in this topology. Because current generators verify only coordination number, neither this charge incompatibility nor potential symmetry mismatches between the building blocks and the net's vertex and edge site symmetries are caught before the structure is written to a CIF file.

White et al. found that most hypothetical MOF databases built from CoRE-derived SBUs exhibit error rates exceeding 40%[12b]; even databases not derived from CoRE SBUs, such as the Wilmer (40%) and Boyd-Woo (21%) libraries, carry substantial error rates. Because D3 demons are structurally complete and valid CIF files, they pass through file-level checks without triggering errors and can only be detected by chemical validation methods such as MOSAEC, MOFChecker, or MOFClassifier. When these structures then serve as training data for ML and generative models, the errors become embedded in the models themselves, creating the feedback loop discussed in **Section 2.2.3**.

**(D4) Expert curation.** A smaller but documented fraction of structural demons is introduced during curation of deposited structures. Although expert review is generally a corrective step, it can introduce errors when, for example, disorder is resolved incorrectly, partial occupancies are rounded to integer values without chemical justification, or protonation states are misassigned. BAKGIF (**Figure 2d**) illustrates this entry point. Inspection of the original deposited CIF file (CCDC 826920) and the publication shows that the crystallographic model is a 3D porous network of $[Ni^{II}_2(H_{0.67}bdt)_3]$ ($H_2bdt$ = 5,5'-(1,4-phenylene)bis(1*H*-tetrazole))[38]. The tetrazole hydrogen atoms on ligated $H_2bdt$ are highly acidic and would normally be fully deprotonated. However, to satisfy charge balance without invoking pore-resident countercations, the authors interpreted the three ligands in two different crystallographic environments as one fully deprotonated $bdt^{-2}$ and two singly deprotonated $Hbdt^{-1}$. The phenyl group of the $bdt^{-2}$ on a crystallographic *mmm* symmetry site shows statistical disorder with partial occupancy. Meanwhile, because $Hbdt^{-1}$ is singly deprotonated, the hydrogen atom on the $Hbdt^{-1}$ at a crystallographic *m*2*m* symmetry site was placed on four symmetry-related nitrogen atoms with 1/4 site occupancy each. Therefore, a more informative way to write this crystal structure is $[Ni_2(bdt)(Hbdt)_2]$, which makes the partial protonation explicit. Considering the highly acid nature of the ligated $H_2bdt$, all ligands might be fully deprotonated $bdt^{2-}$, and the contents of the pore channels, which are treated with SQUEEZE/PLATON and not visible in the CIF, could include charge-balancing cations. An equally plausible alternative is that charge-balancing cations, such as $[N(butyl)_4]^+$ or $H_3O^+$, are present in the solvent-filled pores, giving a fully deprotonated framework $(cation)_2[Ni_2(bdt)_3]$. Neither the deposited CIF nor the publication provides enough information to distinguish between these two scenarios. The BAKGIF case with no partial occupancy information in the CIF file carries a broader lesson. Because the error in the protonation state of the ligand was introduced during curation rather than by an algorithm, it carries an implicit authority that makes it hard to catch automatically: a downstream user has no reason to question the deposited protonation state. Resolving such errors requires going back to the original paper for synthesis and crystallization conditions, showing that even authoritative deposited records cannot always be taken at face value. The frequency of **D3** errors across the full CSD has not been systematically quantified, but even a low rate matters because these errors are inherited by every downstream database that builds on the CSD. **Figure 2e** outlines a decision tree for triaging a given CIF file: checking sequentially for disorder, missing protons or counterions, and oxidation-state anomalies, with each positive finding routed to literature or expert consultation before the structure is accepted as computation-ready.

**Table 1.** Comparison of structural demon detection methods. These methods are complementary; headline accuracy values are not directly comparable because they use different error definitions, benchmarks, and output formats.

| Method | Type | What it checks | Key limitation | Repair? |
|---|---|---|---|---|
| Chen-Manz | Rule-based | Bond order, overlapping atoms (~25% flagged in CoRE MOF 2019 DB) | Sensitive to atomic-radii parameters; can mislabel valid MOFs | No |
| MOFChecker | Rule-based | Geometry + EQeq charges | Over-flags unusual coordination environments | Partial |
| MOSAEC | Rule-based | Metal oxidation states (96% acc., 14,796 structures; 1.9M scanned) | May miss purely geometric distortions | No |
| SETC | ML (GAT) | Error type: H omission, charge, disorder (AUC 0.925-0.949) | Inherits MOSAEC blind spots via input features | No |
| MOF Classifier | ML (PU-CGCNN) | Crystal-likeness score (AUC 0.979) | Global score only; does not identify which atoms or motifs are problematic | No |
| MOF-ChemUnity | Literature | Structure-literature linking | Not a direct validity benchmark | Indirect |
| LitMOF | Literature (LLM) | CIF vs. publication cross-check | Benchmarking still immature | Yes |

**Propagation pathways.** These four entry points are not independent. **D1** demons, if unresolved, propagate through **D2** during post-processing. **D1** and **D2** errors in experimental databases contaminate the SBU libraries from which hypothetical MOFs are built, seeding **D3** demons in hypothetical databases (orange dashed arrows in **Figure 1**). When those databases then serve as training data for generative models (**Section 2.2.3**), the chain extends further: the models may produce new structures that inherit the same chemical implausibilities, feeding them back into the next generation of training data. Recognizing these propagation pathways is essential for designing interventions that address root causes rather than symptoms, which we discuss in **Section 4**.

### 3.2 Rule-Based Validation

The first line of defense against structural demons has been rule-based algorithms that encode specific chemical criteria. Several such methods have been developed, but they do not evaluate exactly the same label or benchmark. Chen and Manz screened the CoRE MOF 2019 database for overlapping atoms, missing hydrogen atoms, and bonding irregularities using bond-order analysis, flagging approximately 25% of the structures as problematic[12a]. Jin et al. formalized and extended MOFChecker[12c], a modular open-source tool that combines geometric checks (overlapping atoms, unreasonable bond lengths) with EQeq charge-based analysis to identify structures with unphysical atomic charges. More recently, White et al. introduced MOSAEC, which assigns metal oxidation states to each metal centre via a bond-valence sum methodology and flags structures in which the result is chemically impossible or highly unlikely[12b]. MOSAEC was manually validated against 14,796 structures from the CoRE database with 96% accuracy and, when applied to 14 major experimental and hypothetical MOF databases containing over 1.9 million structures, revealed structural error rates exceeding 40% in most cases.

These methods should therefore be treated as complementary detectors rather than interchangeable measures of structural validity. MOSAEC targets charge-related inconsistencies but may miss purely geometric distortions; MOFChecker is broader geometrically but sensitive to threshold choices and unusual coordination environments; and the Chen-Manz method emphasizes bonding topology but is sensitive to atomic-radii parameters. Because their labels, priors, and benchmarks differ, headline accuracy values are not directly comparable across methods. Any comparison table should state three things clearly: what label the method predicts, what data it uses, and whether it helps repair the structure.

### 3.3 Machine-Learning Validation

While rule-based methods encode explicit chemical criteria, machine-learning approaches learn proxies for structural validity directly from curated examples. Gibaldi et al. developed Structure Error Type Classification (SETC)[13], a graph attention network trained on over 11,000 MOF structures that were manually inspected and labelled by error type. SETC classifies three major error categories (hydrogen atom omissions, charge-balancing errors, and crystallographic disorder) with Receiver Operating Characteristic-Area Under the Curve (ROC-AUC) values ranging from 0.925 to 0.949. Unlike a binary validator, SETC predicts error type, which is valuable because hydrogen omission, charge imbalance, and disorder require different remediation strategies. The best-performing models use only four node features (atomic number and three oxidation-state descriptors from MOSAEC), meaning that SETC inherits both the strengths and blind spots of the underlying rule-based method. Subgraph explainability analysis further showed that the model frequently identifies the same chemically problematic motifs that a human expert would flag, suggesting that it has learned general chemical principles rather than dataset-specific shortcuts.

Zhao et al. took a more general approach with MOFClassifier[12d], a positive-unlabeled crystal graph convolutional neural network (PU-CGCNN) that learns directly from crystal graph representations without relying on predefined chemical rules. Trained on 13,213 validated CR structures and 25,297 unlabeled structures from the CoRE MOF workflow, MOFClassifier predicts a crystal-likeness score (CLscore) and achieves an AUC of 0.979, surpassing the best combined rule-based value of 0.912 reported in the same study. A key strength of MOFClassifier is the recovery of false negatives: 12.1% of the unlabeled structures were reclassified as CR-like, including well-known frameworks such as Cu-BTC that had been incorrectly flagged by rule-based methods due to their unusual open metal site geometries. MOFClassifier has been integrated into the CoRE MOF DB 2025 v1.0 preparation workflow, where it serves as an additional validation layer alongside MOSAEC and other checks. **Table 1** compares what current methods actually do: some flag local chemical problems, some give a global score, and some help recover information that is missing from the CIF.

Despite these advances, current validation methods still share fundamental limitations. Manual curation, the ultimate ground truth, introduces subjectivity which can introduce new types of ambiguity during curation process. Addressing such blind spot will likely require approaches that go beyond global structure-level classification: graph explainability techniques that localize classifier decisions to individual atoms represent a promising path toward integrating the two approaches (**Section 5**). In some cases, however, the decisive evidence lies outside the CIF itself.

### 3.4 Validation and Repair Using the Original Paper

Refinement details, synthesis conditions, and author discussions of disorder are often absent from the deposited structure file but can be essential for deciding whether an apparent error is real. Two recent efforts address this gap from complementary directions. Pruyn et al. introduced MOF-ChemUnity[39], a structured knowledge graph that links literature-derived insights to crystal structures in the CSD and supports literature-informed AI assistants for retrieval and materials recommendation. Kim et al. describe LitMOF[40], an LLM-driven multi-agent framework that cross-references CIF files against the synthesis descriptions and database entries to validate and repair crystallographic information. Both are best regarded as promising proofs of concept; large-scale independent validation remains to be reported for either system. Even so, they support an important conclusion: some structural demons cannot be resolved from deposited geometry alone and must be judged using information outside the CIF. The BAKGIF case described in **Section 3.1** illustrates the point directly: the curation-introduced protonation error was invisible in the deposited CIF, and resolving whether the discrepancy arose from missing counterions or another source requires consulting the synthesis conditions reported in the primary literature. No single demon-hunting strategy is sufficient. The best workflows will combine chemical rules, ML classifiers, and checks against the original paper when the CIF alone is not enough. The central challenge, then, is not simply finding errors but determining what kind of error is present and how it should be addressed.

## 4. Preventing New Demons: Defense at Three Pipeline Stages

The methods described in **Section 3** detect and classify structural demons that already reside in databases. Prevention requires a different question: how can fewer structural demons be introduced in the first place? Three intervention tiers are especially important. **P1** preserves synthesis and measurement context before it is lost. **P2** ensures that the curation path from raw data to computation-ready structure is auditable and reproducible. **P3** enforces chemical validity at the point of hypothetical structure generation. The logic is cumulative: upstream context reduces ambiguity, auditable curation prevents avoidable translation errors, and validity-by-construction keeps implausible hypothetical structures out of downstream databases.

**(P1) Preserve experimental context.** The most direct way to reduce D1 demons is to ensure that experimental context, such as synthesis conditions, measurement parameters, and sample-preparation details remain machine-readable and linked to the deposited structure after publication. Two recent standardization efforts address the two most common points at which this context is lost. Cheung et al. introduced the Material Preparation Information File (MPIF)[41], a STAR-based, machine-readable format that records reagents, conditions, equipment, and handling procedures at the point of synthesis. Built on the same architectural logic as CIF, MPIF is designed to interoperate with structural and adsorption data files, creating a linked record that begins at the synthesis bench. On the characterization side, Wang et al. demonstrated LFAST[42], a framework coupling a GPT-4o-powered research agent with robotic synthesis and high-throughput PXRD characterization. The associated powder X-ray diffraction information file (.pxrdif) keeps the diffraction pattern together with the instrument parameters, sample-preparation details, and measurement conditions needed to interpret it. Together, MPIF and .pxrdif address the two stages at which experimental context is most commonly lost: before characterization and during translation from measurement to deposited structure. When this context accompanies the CIF into downstream databases, many D1 ambiguities such as missing hydrogen atoms, unresolved counterions, and uncertain protonation states become tractable without returning to the original publication for context.

**(P2) Make curation auditable and reproducible.** Even when experimental context is preserved, converting a raw CIF into a computation-ready model requires chemical decisions (solvent removal, disorder resolution, hydrogen addition, charge assignment) that can themselves introduce D2 demons. The defense at this stage is not to eliminate human judgement but to make each decision traceable and reproducible through consistent software pipelines. CoRE-MOF-Tools integrates CIF collection, solvent removal, validation, charge assignment, topology analysis, and featurization into a single reproducible workflow, and its explicit distinction between CR and NCR structures ensures that ambiguous entries are flagged rather than silently forced into completed models[5d]. Jin et al. extended MOFChecker with automated repair routines for missing hydrogen atoms and counterions, providing rule-based corrections whose inputs and outputs can be inspected at each step[12c]. The SAMOSA protocol monitors metal oxidation states throughout the solvent-removal process. Jablonka et al. addressed benchmarking and featurization with mofdscribe and moffragmentor[43]. The common principle is that every transformation from raw CIF to simulation input is logged and reversible. If a downstream user questions a structural decision, the record of each processing step can be inspected without returning to the original paper. P1 ensures that information needed to make correct curation decisions is available; P2 ensures that the decisions actually made are transparent and correctable.

**(P3) Enforce validity at the generation stage.** For hypothetical databases, the analogous principle is to prevent chemically invalid structures from being generated rather than filtering them afterward. Two complementary strategies operate at different points in the generation pipeline. At the input stage, validated SBU libraries such as HEALED[32] reduce database contamination by ensuring that each building block carries a chemically consistent oxidation state, coordination geometry, and charge before it enters any generator. At the topology selection stage, the symmetry-guided filtering workflow introduced by Darù et al.[31] matches SBU point-group symmetry against RCSR vertex and edge site symmetry requirements, rejecting incompatible combinations before any atomic coordinates are generated. As demonstrated for MOF-841, this filtering reduced fifteen candidate topologies to a single correct assignment (**flu**). Because the output is a short, pre-validated topology list rather than a long combinatorial library, DFT-level optimization becomes tractable for every surviving candidate. Incorporating RCSR symmetry constraints as construction-time filters, rather than relying solely on coordination-number matching (**Section 2.2.3**), would narrow the candidate space substantially and reduce **D3** demons entering hypothetical databases.

These tiers are complementary. **P1** reduces the number of ambiguous records that reach curation by preserving the context needed to resolve them. **P2** makes the curation path traceable and reproducible, preventing avoidable **D2** errors. **P3** restricts hypothetical generation to chemically plausible outcomes at the blueprint stage. Mapped onto **Section 3.1**, **P1** mainly suppresses **D1** entry, **P2** intercepts **D1** and **D2** before they propagate, and **P3** reduces **D3** by enforcing chemical validity before substructures are fully assembled. **D4** remains the hardest class because curation errors often can be resolved only by going back to the original paper.

## 5. Outlook

This review has tracked structural demons from their four entry points, experimental characterization, automated post-processing, expert-curation, and *in silico* generation, through the validators that detect them and the infrastructure that can prevent them.

Three conclusions follow. First, detection is no longer the central bottleneck. Rule-based validators and ML classifiers are ready for routine use in curation workflows. Systems that check the original paper are promising, but they still need broader validation. The next technical advance should be elevating them to address finer details: methods that identify which specific atoms or motifs drive a failed classification would help connect ML outputs to chemically interpretable repair actions.

Second, hypothetical databases need stronger validity benchmarks. Generative models should not be judged only by novelty, diversity, or geometric realism, but also by whether their outputs satisfy the same chemical criteria enforced by MOSAEC, MOFClassifier, and SETC. Validated SBU libraries remain an essential foundation, and the broader opportunity is to use RCSR labelled quotient-graph, vertex symmetry, and space-group constraints as construction-time filters rather than post-hoc descriptors.

Third, the path from experiment to database should be treated as a shared resource that is updated in a controlled and transparent way. New validators, repaired structures, revised metadata, and updated property calculations should propagate across database releases so that improvements accumulate rather than remain isolated. The BAKGIF case also shows why literature-grounded validation must remain part of this infrastructure: some errors reside in authoritative deposited records and cannot be diagnosed with geometry alone. An open-access, community-governed structure repository for reticular materials, analogous to what the Protein Data Bank provides for structural biology, would remove this barrier and allow the pipeline improvements described here to benefit the field as a whole.

Structural demons are a bigger problem than a few bad entries in a database. They are a systems problem that spans measurement, curation, generation, and model training. The tools to find these demons now exist, but hunting demons one at a time is not the endgame. The next step is to connect generation, measurement, curation, and validation so that errors are harder to introduce, easier to spot, and removed before they spread. The reward is not only a cleaner archive but a more reliable path from digital design and discovery to experimental realization.

## Acknowledgement

This work was supported by the National Research Foundation of Korea (NRF) Grant funded by the Korean government (RS-2024-00421195, RS-2024-00449431).

Biographical Sketch

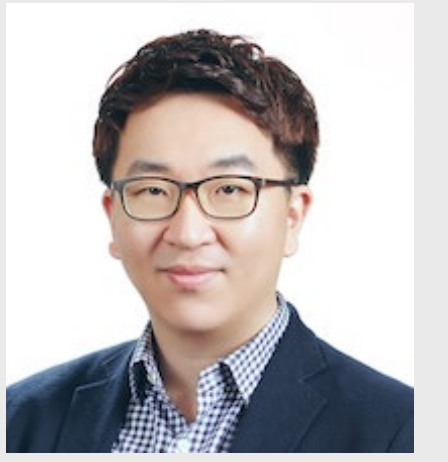

Yongchul G. Chung is a Professor of Chemical Engineering at Pusan National University, Busan, Korea. He received Ph.D. in Chemical Engineering from Case Western Reserve University (Advisor: Daniel J. Lacks), followed by postdoctoral research with Randall Q. Snurr at Northwestern University. He led the development of the CoRE MOF Database. His current research focuses on molecular simulation of nanoporous materials, force field development, and AI-driven materials and chemical discovery, with an emphasis on building reliable data infrastructure for digital reticular chemistry.

Biographical Sketch

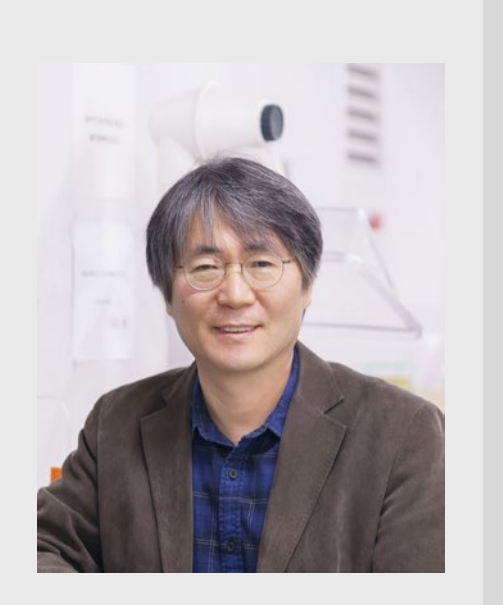

Myoung Soo Lah is an Emeritus Professor of Chemistry at the Ulsan National Institute of Science and Technology (UNIST), Korea. He received his Ph.D. from the University of Michigan and served on the faculty of UNIST, where his research centered on the structural chemistry of metal–organic frameworks (MOFs) and coordination polymers. His expertise in the design, synthesis, and characterization of MOFs has contributed to the development and structural elucidation of MOF systems deposited in the Cambridge Structural Database.